\documentclass[prl,twocolumn,aps,superscriptaddress,showpacs,amsmath]{revtex4}
\usepackage{graphicx,bm}
\begin{document}

\title{Interchain-Frustration-Induced Metallic State
 in Quasi-One-Dimensional Mott Insulators}

\author{M.\ Tsuchiizu}
 
\affiliation{Department of Physics, Nagoya University, 
Nagoya 464-8602, Japan}

\author{Y.\ Suzumura}
 
\affiliation{Department of Physics, Nagoya University, 
Nagoya 464-8602, Japan}

\author{C.\ Bourbonnais}

\affiliation{D\'{e}partement de Physique, Universit\'{e} de Sherbrooke, 
 Sherbrooke, Qu\'{e}bec, Canada J1K-2R1}

\date{September 28, 2007}

\begin{abstract}
The mechanism that drives a metal-insulator transition
 in an undoped quasi-one-dimensional Mott insulator
 is examined in the framework of the Hubbard model
 with two different hoppings $t_{\perp1}$ and $t_{\perp2}$
 between nearest-neighbor chains.
By applying an $N_\perp$-chain renormalization group method
  at the two-loop level, 
 we show how a metallic state emerges
  when both $t_{\perp1}$ and $t_{\perp2}$ exceed critical values. 
In the metallic phase, the quasiparticle weight becomes finite
 and  develops a strong momentum dependence.
We discuss the temperature dependence of the resistivity
 and the impact of our theory in the understanding
 of recent experiments on half-filled molecular conductors.
\end{abstract}

 
\maketitle

The remarkable properties of 
  strongly correlated systems near   a metal to Mott  insulator (MI) transition
  stand out as one of the    richest  parts    of  the physics  of
  strongly correlated systems \cite{Imada1998}. 
This takes on particular importance in low dimensional  materials at
  half-filling, and especially in one dimension, 
where spin  and  charge  excitations are well  known to be invariably
  decoupled, an effect that receives experimental confirmation in
 the one-dimensional (1D) oxide material
  SrCuO$_2$ \cite{Kim2006}. 
How this picture  modifies when the  hopping of electrons in more than
 one spatial direction  progressively   grows
and a  higher dimensional metallic ground state becomes possible
 is a key issue  that remains   poorly understood.   
  This problem finds concrete applications in organic molecular
  compounds,  which constitute  very close realizations  
of  1D   systems. This is the case notably of 
  (TTM-TTP)I$_3$ and (DMTSA)BF$_4$ 
  which, as genuine half-filled  band materials
  due to the monovalent  anions I$_3^-$ and BF$_4^-$,
 see their Mott insulating state being gradually
  suppressed under pressure \cite{Mori1997,Mori2004}. 
Quite recently, resistivity measurements on
   (TTM-TTP)I$_3$ at   high pressure
 revealed that the temperature scale for the Mott insulating  behavior 
   is suppressed by an order of magnitude down to 
 $T_\mathrm{MI}\approx $ 20 K at $P =$ 8 GPa of pressure,
   with a  metallic ground state expected
  to  occur   above 10 GPa \cite{Yasuzuka}.
 It is the objective of this letter to propose a theoretical description 
 of this transition.

 Bosonization, renormalization group (RG) approaches to  the 1D Hubbard
 model in weak coupling and its  exact solution from  the Bethe ansatz  
 \cite{Dzyaloshinskii72,Bourbonnais2003,Giamarchi_book},
 show that electron-electron umklapp-scattering processes    
    are a key ingredient that promotes  the existence of a Mott
 insulating ground state at half-filling. 
The difficulty underlying the mechanism of the MI transition in the
 quasi-1D case resides in the fact that the decoupling of spin and
 charge excitations  on the metallic  side of the transition  at higher
 dimension 
does not occur,
 namely  when   Fermi-liquid quasiparticles excitations  appear.
This issue has been addressed theoretically
  by an RPA treatment of 
     interchain hopping \cite{Essler2002},
 in which  its feedback effect   on
  the Mott gap is neglected for the self-energy
  of the one-particle Green's function.
  The RPA results   shows  the existence of a Fermi-liquid metallic state
  with electron and hole Fermi-surface pockets when the interchain 
  hopping exceeds a critical value.
The important role of 
this feedback effect has been pointed out  recently from the use  of dynamical mean-field theory,
  extended   to include the influence of 1D fluctuations 
  (the chain-DMFT) \cite{Giamarchi2001} 
 -- an approach expected to be workable in the strong-coupling regime.
In the weak-coupling regime, however, the electron-electron umklapp scattering 
   and the characteristics of its coupling to nesting of  the whole  Fermi surface,  play a key role 
    \cite{Dzyaloshinskii72,Bourbonnais2003,Giamarchi_book},  both for
 the Mott insulating state  and for the possibility of the MI transition
 that emerges in the quasi-1D case.  
Transverse dispersion 
  which has a tendency towards the realization of the
 deconfined metallic states
  \cite{Bourbonnais2003,Giamarchi_book,Giamarchi2001} 
   is in general overlooked  in the calculation of one-particle Green's function from mean-field like approaches. 
    The two-loop RG approach    avoids  such an approximation  and treats the Fermi-surface nesting  conditions properly. These 
 can be strongly   altered in the quasi-1D case,
especially in the presence of frustration 
in the electron kinetics.

In this Letter, we apply an  $N_\perp$-chain RG approach  
  to  quasi-1D  half-filled system as developed recently at
  the two-loop level \cite{Tsuchiizu2006,Tsuchiizu2006b}. The  momentum
  dependence for 
   nesting of  the whole Fermi surface 
  and  that for   couplings, including   umklapp scattering  
  \cite{Duprat2001,Doucot2003,Rohe06}, 
are  taken into account in a systematic way
in the calculation of one-particle Green's function  and four-point
  vertices  \cite{Tsuchiizu2006}.
 The combined impact of correlations and  nesting frustration
  induced by interchain hopping on the  transverse momentum dependence of the 
 quasiparticle weight is obtained in  
     weak-coupling regime.

\begin{figure}[t]
\includegraphics[width=7cm]{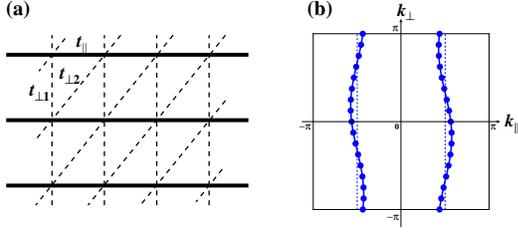}
\caption{
(Color online).
(a) Lattice geometry of the present model.
(b) The corresponding Fermi surface where
the case for $N_\perp =16$  is shown.
}
\label{fig:model}
\end{figure}

We consider the quasi-1D half-filled Hubbard model 
 on  an anisotropic 
triangular lattice  [Fig.\ \ref{fig:model}(a)], with
  the transfer energies
  $t_\parallel \gg |t_{\perp 1}|,|t_{\perp 2}|$ 
  ($t_\parallel$ is  the energy along chains and 
  $t_{\perp1}$  and  $t_{\perp2}$  
  are those between chains).
Our Hamiltonian, with the on site Coulomb   repulsion $U$
is given by 
%====================================================================
\begin{eqnarray}
H &=& 
-t_\parallel \sum_{j,\ell,s} 
\left(c_{j,\ell,s}^\dagger c_{j+1,\ell,s}^{}+\mathrm{H.c.} \right)
\nonumber \\ && {}
-t_{\perp 1} \sum_{j,\ell,s} 
\left(c_{j,\ell,s}^\dagger c_{j,\ell+1,s}^{}+\mathrm{H.c.} \right)
\nonumber \\ && {}
-t_{\perp 2}\sum_{j,\ell,s} 
\left(c_{j,\ell,s}^\dagger c_{j+1,\ell+1,s}^{}+\mathrm{H.c.} \right)
\nonumber \\ && {}
+ U \sum_{j,\ell} n_{j,\ell,\uparrow} n_{j,\ell,\downarrow}
-\mu \sum_{j,\ell,s}  c_{j,\ell,s}^\dagger c_{j,\ell,s}^{}.
\label{eq:model}
\end{eqnarray}
%====================================================================
The operator  $c_{j,\ell,s}$ denotes the annihilation  of an electron 
  on the $j$th site in the $\ell$th chain with spin $s$, and 
 $n_{j,\ell,s}=c_{j,\ell,s}^\dagger c_{j,\ell,s}^{}-\frac{1}{2}$.
Here  $\ell=1,\cdots,N_\perp$ is the chain index.
We take the continuum limit along the chain direction, whereas
 in the transverse direction we consider a finite system
 having an even number of chains $N_\perp$ with  
 the boundary condition  $c_{j,N_\perp+1,s}=c_{j,1,s}$.   
By applying a  Fourier transform,
the kinetic term can be rewritten as
 $H_0 = \sum_{\bm k,s}
\varepsilon(\bm k) \,
c_s^\dagger (\bm k) \,c_s^{} (\bm k)$, 
where $\bm k \equiv (k_\parallel,k_\perp)$ and
   the energy dispersion is given by
$\varepsilon(\bm k) =
  -2t_\parallel \cos k_\parallel 
  -2t_{\perp 1} \cos k_\perp
  -2t_{\perp 2} \cos (k_\parallel + k_\perp) - \mu$.
For  $|t_{\perp i}| \ll t_\parallel$,
  the right- and left-moving electrons in the 1D case are well-defined 
  due to  an open Fermi surface [Fig.\ \ref{fig:model}(b)] and 
  the shape of the Fermi surface can be parametrized  by the 
    transverse momentum  $k_\perp$ \cite{Tsuchiizu2006}.
To lowest order in the  interchain hoppings 
  $t_{\perp 1}$ and $t_{\perp 2}$,
the Fermi surfaces for right $(+)$ and left $(-)$ moving electrons, 
  as a function of $k_\perp$,
  are given by 
$k_F^\pm(k_\perp)
= \pm  (\pi/2) 
\pm (t_{\perp 1}/t_\parallel) \cos k_\perp
- (t_{\perp 2}/t_\parallel) \sin k_\perp$
and  the chemical potential is   $\mu=O(t_{\perp i}^2)$.
By considering the weak-interacting case 
 and neglecting the $k_\perp$ dependence of the velocity,
    the linearized dispersion used in  the RG method takes the simple form:
$\varepsilon_p (\bm k)= p v[k_\parallel -k_F^p(k_\perp)]$
with $v=2t_\parallel$ and $p=\pm$. 

%====================================================================
%======================================================================

We follow the formulation of the two-loop RG 
  method of Ref.\ \cite{Tsuchiizu2006}
and first introduce the $g$-ology coupling constants
 [see Eq.\ (2.8) in Ref.\ \cite{Tsuchiizu2006}]:,
 namely   the backward ($g_{1\perp}$), forward  ($g_{2\perp}$),
 and umklapp  ($g_{3\perp}$) scatterings with  opposite spins, and 
   the forward scattering  ($g_\parallel$)  and umklapp
  ($g_{3\parallel}$) scatterings for parallel spins.
The coupling constants are 
   renormalized differently, developing an external -- transverse -- momenta 
  dependence in the vertex corresponding to a  patch-index  dependence;
 that is,  $g_\nu \to g_{\nu(q_\perp,k_{\perp 1},k_{\perp 2})}$,
  where $k_{\perp 1}$ and $k_{\perp 2}$ are the transverse momenta
  for the right-going fermions,  and $q_\perp$ is the 
  momentum transfer \cite{Tsuchiizu2006}.  
The magnitude  of the  initial couplings are
  given by   
  $g_{1\perp}=g_{2\perp}=g_{3\perp}=U$ and 
  $g_{\parallel}=g_{3\parallel}=0$.
The RG equations   are derived by 
 scaling the bandwidth cutoff $\Lambda(\simeq 2\pi t_\parallel )$  as
   $\Lambda_l= \Lambda e^{-l}$,  where $l$ is the scaling parameter 
  \cite{Tsuchiizu2006}.
The explicit forms of the two-loop RG equations 
  for all   coupling constants 
in the case  $t_{\perp2}=0$ are 
 given in Ref.\ \cite{Tsuchiizu2006}.
We solve here these  two-loop RG equations numerically for 
 a system with $N_\perp=16$.
For even $N_\perp$, the solution of the RG flows
  indicates the existence of a finite spin gap in the low-energy limit, 
   which is  expected to vanish 
  in the infinite $N_\perp$ limit.  
The characteristic scale $l_{N_\perp}$ above which
  finite size effect would appear can be 
  roughly estimated  to be 
  $l_{N_\perp} \approx 
   \ln [\Lambda/|t_{\perp i} \sin (2\pi/N_\perp)|]$.
For   $N_\perp=16$ 
and $t_{\perp 1}/t_\parallel=0.1$ 
   this gives   $l_{N_\perp}\approx 5$, 
  which is sufficient to obtain the MI transition 
 without  finite size effects 
for the present choice of parameters $U/t_\parallel=2$ or $1.5$.

For the bipartite lattice ($t_{\perp2}=0$),
   the system remains always insulating 
  even for large $t_{\perp1}$ \cite{Tsuchiizu2006}.
This is due to   
 perfect nesting condition for the Fermi surface.
In this case the interchain hopping $t_{\perp1}$ is relevant and becomes 
   large under scaling procedure, and 
  some umklapp
  couplings become small. However, 
 a macroscopic number of these
 couplings  remains relevant
  due to   perfect nesting at   vector 
   $\bm Q=(\pi,\pi)$.
In the case of large $t_{\perp2}$,
   these relevant umklapp 
 couplings
 are strongly reduced 
 and all umklapp terms
 remain weak.
The charge gap then collapses 
 and  a metallic state 
 with a Fermi surface emerges as a consequence of 
    nesting deviations   that are introduced
by large interchain frustration.

\begin{figure}[b]
\includegraphics[width=8.5cm]{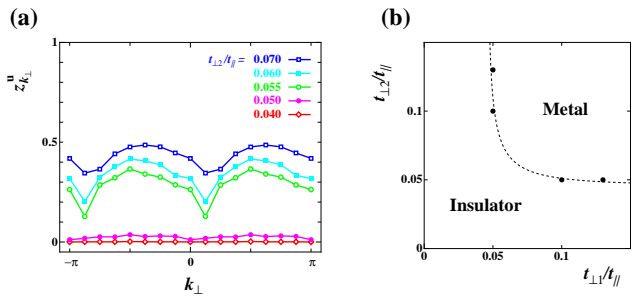}
\caption{
(Color online).
(a) Umklapp-scattering contribution of the
   quasiparticle weight 
   $z_{k_\perp}^{\mathrm{u}}$  
 for $U/t_\parallel=2$ and
  $t_{\perp1}/t_\parallel=0.1$, 
 with several $t_{\perp2}$.
(b) Ground-state phase diagram 
  on the plane of $t_{\perp1}/t_\parallel$ and $t_{\perp2}/t_\parallel$.
The dotted line denotes the condition of Eq.\ (\ref{eq:boundary}).
}
\label{fig:phasediagram}
\end{figure}

One-particle properties are best studied  from the 
 quasiparticle  weight   $z_{k_\perp}$.
In the RG method at the two-loop level \cite{Tsuchiizu2006}, 
  the $k_\perp$ dependence of the 
  self-energy is taken into account in a non perturbative way.  This contrasts  with the chain-DMFT,  which treats 
  the one-particle self-energy, as a one-chain
   $k_\perp$-independent quantity  of 
   the one-particle Green's function  
  \cite{Giamarchi2001}.    In the RG formalism
 the transverse momentum dependence \cite{Tsuchiizu2006} 
  is taken into account  explicitly, and  
    the quasiparticle weight of the one-particle Green's function
   takes the form
      $z_{k_\perp}\equiv z_{k_\perp}^{n} z_{k_\perp}^{u}$,
  where $z_{k_\perp}^{n}$ and $z_{k_\perp}^{u}$
  are the contributions  
 coming from the normal and umklapp parts of the scattering,
  respectively.
Here, 
  one can  focus   on the umklapp contribution
 $z_{k_\perp}^{u}$, 
  since $z_{k_\perp}^{n}$ remains finite for
 $l<l_{N_\perp}$  in the metallic state.
The explicit form of $z_{k_\perp}^{u}$ reads
\cite{Tsuchiizu2006} 
%====================================================================
\begin{eqnarray}
&& \hspace*{-.5cm}
z_{k_{\!\perp}}^{u} 
\! = \!
\exp
\left[
-\frac{1}{2N_\perp^2} \sum_{q\!_\perp,k'_{\!\perp}} \int
dl \,
 G_{\Sigma u(q_{\!\perp},k_{\!\perp},k'_{\!\perp})}^2 \,
  J^{u}_{1(q_{\!\perp},k_{\!\perp},k'_{\!\perp})}
\right],
\nonumber \\
\label{eq:z}
\\
&& \hspace*{-.5cm}
G_{\Sigma u(q_\perp,k_\perp,k'_\perp)}^2  \equiv
  G_{c(q_\perp,k_\perp,k'_\perp)}^2
+ G_{c(\pi-q_\perp+k_\perp+k'_\perp,k_\perp,k'_\perp)}^2
\nonumber \\ && {}\qquad\qquad
- G_{c(q_\perp,k_\perp,k'_\perp)}  G_{c(\pi-q_\perp+k_\perp+k'_\perp,
      k_\perp,k'_\perp)},
\end{eqnarray}
%====================================================================
where $G_{c(q_\perp,k_\perp,k'_\perp)}
  \equiv g_{3\perp(q_\perp,k_\perp,\pi-k'_\perp)}/(2\pi v)$.
The quantity
  $J^{u}_{1(q_\perp,k_\perp,k'_\perp)}$ is a
  cutoff function depending on the transverse dispersion 
 \cite{Tsuchiizu2006}
  given by 
  $J^{u}_{1(q_\perp,k_\perp,k'_\perp)} =1$ for
  $|A_{q_\perp,k_\perp,k'_\perp}'|\ll \Lambda$ and 
   $J^{u}_{1(q_\perp,k_\perp,k'_\perp)} \approx 0$ for
  $|A_{q_\perp,k_\perp,k'_\perp}'|\gg \Lambda$ where
$A_{q_\perp,k_\perp,k'_\perp}' \equiv
  2t_{\perp 1} [\cos k _\perp + \cos (k_\perp-q_\perp)
                - \cos k'_\perp - \cos (k'_\perp-q_\perp)]
- 2t_{\perp 2} [\sin k_\perp - \sin (k_\perp-q_\perp)
                + \sin k'_\perp - \sin (k'_\perp-q_\perp)]$.
The $k_\perp$ dependence of 
 the umklapp-scattering contribution to $z_{k_\perp}$
  for several values of $t_{\perp2}$
   at  fixed $t_{\perp1}/t_\parallel=0.1$
   is shown in  Fig.\ \ref{fig:phasediagram} (a).
For weak frustration $t_{\perp2}(<t_{\perp2}^{\mathrm{c}})$, this quantity 
is small    showing the existence of an insulating phase 
with no  low-energy quasiparticles.
On the other hand, 
  for strong frustration $t_{\perp2}(>t_{\perp2}^{\mathrm{c}})$,
 it takes sizable values  and shows  a strong $k_\perp$ dependence.
In the proximity of the insulating phase,
 the quantity $z_{k_\perp}^{u}$  presents a broad maximum
 around $k_\perp \approx \pm \pi /2$, 
which behavior implies the emergence of
 sections of Fermi surface 
 \cite{Essler2002}, or  ``cold'' regions
 around $\bm k \approx  (k_F^p(\pm\pi/2),\pm\pi/2)$
 \cite{Duprat2001}. 
 The dips or ``hot spots'' in the quasiparticle weight correspond to
 regions of the Fermi surface where the nesting, 
 within  $2\pi/N_\perp$ accuracy, remain
  the most favorable.
In the  perfectly nested case    no  Fermi surface is found for  
       any value of $U$.  The overall profile is then intrinsically linked to
  the momentum dependent nesting properties 
  over the whole Fermi surface \cite{Rohe06}. These are not taken 
  into account in previous analyses 
   \cite{Giamarchi2001,Essler2002},
   which reach different conclusions in the weak-coupling regime.      

%====================================================================
%======================================================================

The boundary for the metal-insulator transition 
  can be determined analytically
  by noting that
  the metallic phase  of the RG analysis  is linked to the irrelevance 
 of umklapp scattering. 
The metallic phase boundary  can thus be obtained when 
  the energy scale of  imperfect nesting  
  for  umklapp scatterings
  becomes comparable to the energy scale 
  $\Delta_\rho^{\mathrm{1D}}$ [$\propto \sqrt{t_\parallel U} \exp(-2\pi t/U)$]
 of  the 1D  Mott gap.
By noting that the nesting vector of the particle-hole loop,
 which couples to umklapp,
 is given by
  $\bm Q = (\pi, \pi \pm 2\alpha )$ with $\tan \alpha=t_{\perp2}/t_{\perp1}$,
the degree of the imperfect nesting (the amplitude 
  of the quantity $A'_{\pi\pm 2\alpha,k_\perp,k'_\perp}$) becomes 
$\sqrt{t_{\perp 1}^2 + t_{\perp2}^2}  \sin 2 \alpha$.
The   phase boundary  for the MI transition  is
 then determined by the condition    
%====================================================================
\begin{equation}
\frac{t_{\perp1}t_{\perp2}}{\sqrt{t_{\perp 1}^2 + t_{\perp2}^2}}
= c  \Delta_\rho^{\mathrm{1D}},
\label{eq:boundary}
\end{equation}
%====================================================================
where  $c $ is a numerical constant being of the order of unity.
The small difference between the numerical results and 
  the above analytical expression comes from  
  the renormalization of   interchain hopping due to 
  the normal scattering processes. The Mott gap is
  also renormalized by   interchain hopping.
The ground-state phase diagram in the $(t_{\perp1}/t_\parallel, t_{\perp2}/t_\parallel)$ plane   is shown in 
 Fig.\ \ref{fig:phasediagram}(b), where 
 the dotted line denotes Eq.\ (\ref{eq:boundary}).
  The ambiguities in the  cutoff functions of the RG
  \cite{Doucot2003,Tsuchiizu2006},
      prevent  us  from obtaining a precise location of the phase
      boundary.

The Mott transition  has also been 
  addressed in two-dimensional Hubbard model 
  on an anisotropic triangular lattice 
 with   nearest-neighbor hopping $t$ and 
  next-nearest neighbor hopping $t'$
   \cite{Imada_Kyung}.   
The present model (\ref{eq:model}) can be connected to the
  two-dimensional Hubbard model
  by taking 
  $t_{\parallel}\to t$, $t_{\perp1}\to t$, and
  $t_{\perp2}\to t'$. 
While our approach  is restricted to the small 
 interchain hopping,
   the    metal-insulator transition obtained here 
for finite frustration  is consistent with the numerical results
  in two dimensions.

\begin{figure}[t]
\includegraphics[width=8.5cm]{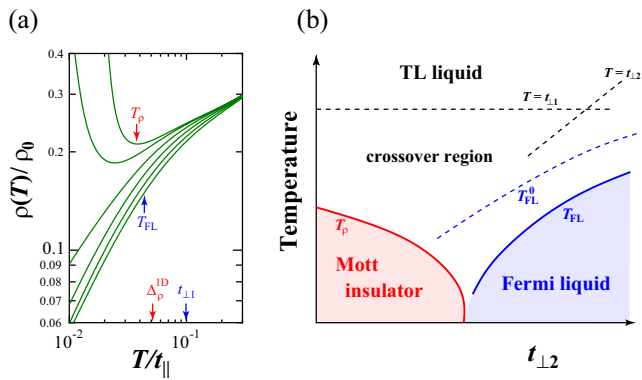}
\caption{
(Color online).
(a)Temperature dependence of the resistivity 
  for $U/t_{\parallel}=1.5$ and $t_{\perp1}/t_{\parallel}=0.1$,
  with fixed $t_{\perp2}/t_\parallel=$ 0, 0.02. 0.04, 0.06, 0.08,
  0.10 (from top to bottom), and $\rho_0=\rho(\Lambda)$.
$T_\rho$
  is the characteristic temperature
  of the Mott insulator, at which the resistivity takes a minimum.
$T_{\mathrm{FL}}$ is the crossover temperature
  to the metallic Fermi liquid, where the power of $\rho(T)$ exhibits
  a change (guided by the eyes).
(b) Schematic illustration of the $t_{\perp2}$-$T$ phase diagram.
 $T_{\mathrm{FL}}^0$ is
  the bare temperature scale for imperfect nesting 
  [left of Eq.\ (\ref{eq:boundary})], which is reduced to 
  $T_{\mathrm{FL}}$ due to the correlation effects.
}
\label{fig:resis}
\end{figure}

From the solution of the scaling flows of the umklapp scatterings,
  the temperature dependence of the resistivity can be qualitatively
 calculated 
 from the  memory function approach
  combined with the RG method
  by using $l=\ln (\Lambda/T)$ \cite{Giamarchi_book,Giamarchi1991}.
 By extending the approach to the quasi-1D case,
   the perturbative expression of the conductivity reads 
$\rho(T) \propto  N_\perp^{-3}  \sum_{q_\perp,k_\perp,k'_\perp} 
G_{c(q_\perp,k_\perp,k'_\perp)}^2(l)  \, e^{-l}$.
While this  formula is not valid for large $t_{\perp i}/T$
  \cite{Giamarchi1991}, 
 it will depict the qualitative  temperature dependence
 of resistivity. 
Typical behaviors of the resistivity 
 for  $\Delta_\rho^{\mathrm{1D}}< t_{\perp 1}$,
 obtained 
  from this formula, are shown in Fig.\ \ref{fig:resis}(a).
At high temperature ($T>t_{\perp 1}$, $t_{\perp2}$),
  the effects of interchain kinetics are masked due to the 
  thermal fluctuations where the Tomonaga-Luttinger (TL) liquid behavior 
   is reproduced.
At low temperature, the insulating behavior can be seen 
  for small $t_{\perp2}$, 
 while the metallic behavior is found for strong
  frustration (large $t_{\perp2}$).
A finite-temperature phase diagram, 
 as a function of $t_{\perp2}$, 
can be obtained from this behavior,
  as  shown schematically 
 in Fig.\ \ref{fig:resis}(b).
At high temperature, 
the TL liquid state is realized,
 which is followed 
    in the crossover region by the development of a 
  Fermi surface.
 The effect of the frustration 
  is yet masked by thermal fluctuations,  i.e.,
  this region  can be described effectively by a nested Fermi surface.  
At further low temperature, the state moves to the Mott insulator 
  or the Fermi liquid (FL) state   depending on $t_{\perp2}$.
In the FL state, the effect of $t_{\perp2}$ becomes prominent and
 the nesting conditions of the full Fermi surface 
  are altered.
It can also be seen from  Fig.\ \ref{fig:resis}(a) that 
  the characteristic temperature $T_{\mathrm{FL}}$ at which the 
   crossover   to the metallic FL state  takes place 
   increases with $t_{\perp 2}$.
For $\Delta_\rho^{\mathrm{1D}} >  t_{\perp 1}$,
 on the other hand,  a direct crossover from the TL liquid
  state to the Mott insulator would be seen, and neither the FL 
  state nor the crossover region appears in the phase diagram.

%====================================================================
%======================================================================

We would like to  discuss here the impact of our results on the understanding 
of  the phase diagram of  the quasi-1D molecular compound
 (TTM-TTP)I$_3$ under  pressure
  \cite{Yasuzuka}. 
The extended H\"uckel 
  calculations \cite{Mori1994}
   indicate that 
there are two kinds of interchain
   transfer integrals,  namely  $t_{\perp1}\approx 9$ meV, 
and $t_{\perp 2}\approx 6$ meV, 
 which are  small compared to $t_\parallel\approx 260$ meV.
 This  emphasizes the pronounced hopping frustration of this quasi-1D compound.
The Coulomb repulsion   between electrons 
  on the same molecular orbital of TTM-TTP is weak,
  and  the estimated magnitude,  $U \approx$ 0.57 eV \cite{Mori1997},
  leads to the magnitude 
  $U/t_\parallel \approx 2.2$ 
    and then the small 1D Mott gap  
  $\Delta_\rho^{\mathrm{1D}}/t_\parallel \approx 0.24$. 
From these figures, 
  the energy scale for  imperfect nesting 
  [left-hand side of Eq.\ (\ref{eq:boundary})] turns out to be 
  about 5 meV. As for  the magnitude 
  of the charge gap 
  at ambient pressure, we obtain 
  $\Delta_\rho^{\mathrm{1D}} \approx 60$ meV,  
 in fair agreement with  the measured 
  activation energy of   resistivity 
  at ambient pressure \cite{Mori1997}.
The bandwidth goes up under pressure, which decreases  
 the ratio  $U/t_\parallel$ and in turn the charge  gap. 
  To reduce the gap down to the
      scale  of  imperfect nesting,
 one has roughly to double   the hopping amplitudes. 
This  represents a reasonable increase  of band parameters under 8 GPa
of pressure,
and is consistent with the existence of a MI transition
 in (TTM-TTP)I$_3$  \cite{Yasuzuka}.  
  Finally, we note that at ambient pressure, below
  $T_{\mathrm{c}}\simeq$ 120 K, this compound develops  a nonmagnetic
  state (likely of the spin-Peierls type), which comes with an
  \textit{intra-molecular} charge disproportionation  \cite{Mori2004}.
However, since the resistivity already shows
   an insulating behavior above $T_{\mathrm{c}}$,
the present system  can be indeed  considered as  a 
   Mott insulator rather than a charge-ordered insulator, and
the charge disproportionation 
can be accounted for by dimerization of the tilted TTM-TTP molecules.

In summary, we have examined the effect of  interchain frustration 
   on  the half-filled quasi-1D Hubbard chains
  by applying an $N_\perp$-chain two-loop RG method.
The triangular lattice  geometry of the system 
 is found to be a key factor in the stability of the 
  Mott insulating state  and whenever the alteration 
of nesting conditions due to frustration 
  in the transverse hoppings reaches some threshold 
a metallic state is restored.
Our results find  direct application to  the  description of 
 the frustrated quasi-1D half-filled compounds.

One of the authors (M.T.) thanks S.\ Yasuzuka, T.\ Kawamoto,
  T.\ Mori, A.\ L\"auchli, T.\ Giamarchi, and C.\ Berthod
  for valuable discussions.
This work was supported in part by a Grant-in-Aid for
Scientific Research on Priority Areas of Molecular Conductors
(No.\ 15073213) from the Ministry of Education, Science, Sports, and Culture,
Japan.

\end{document}